# An Optical Demonstration of Fractal Geometry

B C Scannell[†], B Van Dusen and R P Taylor
Department of Physics
University of Oregon
1274 University of Oregon
Eugene, Or 97403-1274
[†]E-mail: billy@physics.uoregon.edu

**Abstract**

We have built a Sinai cube to illustrate and investigate the scaling properties that result by iterating chaotic trajectories into a well ordered system. We allow red, green and blue light to reflect off a mirrored sphere, which is contained in an otherwise, closed mirrored cube. The resulting images are modeled by ray tracing procedures and both sets of images undergo fractal analysis. We offer this as a novel demonstration of fractal geometry, utilizing the aesthetic appeal of these images to motivate an intuitive understanding of the resulting scaling plots and associated fractal dimensions.

It is rather simple to create a fractal object such as the Sierpinski triangle by using a mathematical prescription or an iterated function. Since fractal geometry offers a useful description of a wide range of physical phenomena, It is not surprising that a physical process such as light rays scattering off of curved surfaces would produce similarly complex images. Intrigued by the apparatus described by Sweet *et al.*,[1] and others [2], we first built a light scattering system consisting of four reflective spheres stacked in a pyramid formation. Expanding on this we built a "Sinai Cube" [3]which offers an analog of an infinite set of stacked spheres. Using front surface mirrors we construct a cube with openings at the upper corners. We suspend a reflective sphere from the top mirror creating the Sinai cube. Three of the top corners are illuminated with colored light, and in the fourth corner we place a camera to capture the image created by the lights repeated reflection off of the center sphere. Figure 1 shows both systems.

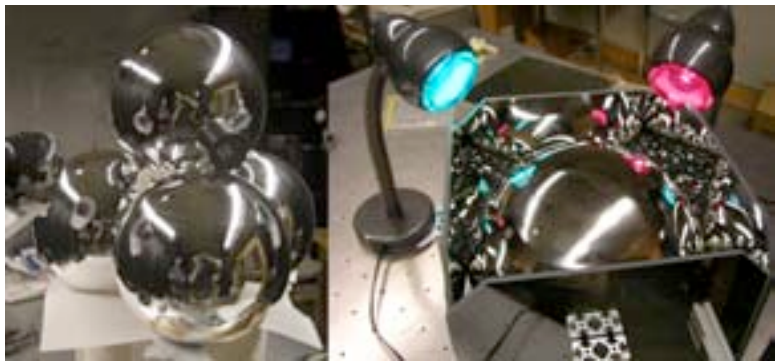

**Figure 1**: *The two experimental apparatuses are pictured above. On the left is a recreation of the laboratory model presented by Sweet et al. On the right is our Sinai cube shown here without the top mirror in place and with a Sinai diffuser whose diameter is approximately equal to the wall width.*

We use ray tracing techniques [4] to model both the stacked sphere and Sinai cube images with. Figure 2 shows the image obtained from the bottom opening of the stacked sphere configuration (top middle) as well as the ray tracing model (directly below). We also show the experimental (top right) and modeled (directly below) Sinai cube in which the diffuser diameter is roughly 1/3 the width between the walls. A box counting analysis was performed on both systems and in each case there was good agreement between the simulated and actual images. For the stacked spheres we found $D=1.6$ in agreement with Sweet *et.al*., similar to the analytic value of the Sierpinski triangle $D =1.58$. In contrast the Sinai cube has significantly higher value of $D = 1.8$. The bottom left corner of Figure 2 we show the resulting scaling plots. Since $D$ is the gradient of the scaling plot it is essentially quantifying the ratios of coarse to fine structure, such that fine structure plays a more dominant role in the high $D$ pattern than the corresponding low $D$ pattern. Consequently the high $D$ pattern appears more visually complex as can be seen by visual comparison of the fractal images generated by the Sinai cube and the stacked spheres.

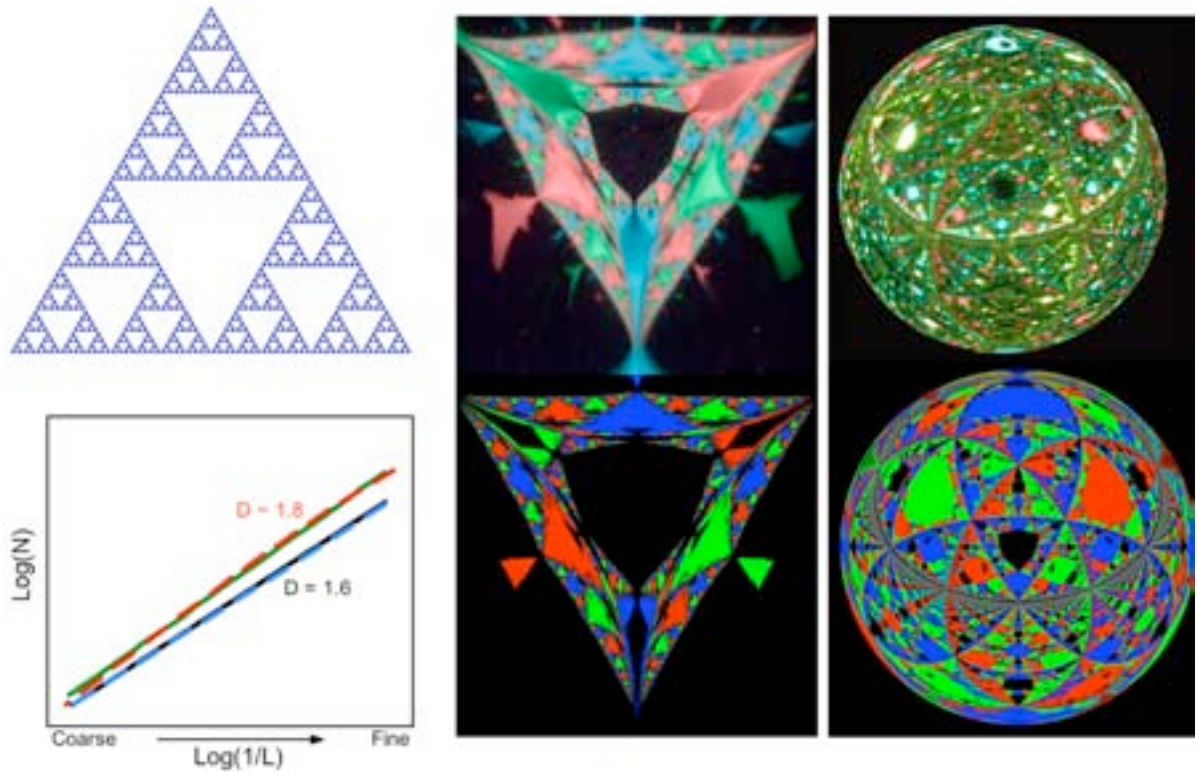

**Figure 2**: *Clockwise from top left: Sierpinski Triangle, image of the stacked sphere configuration, image of the Sinai cube, ray tracing of the Sinai cube, ray tracing of the stacked spheres, scaling plots of the box counting analysis of the Sierpinski triangle (lower dashed blue) the stacked spheres (lower solid black), and the Sinai cube image (upper dashed red) and ray tracing (upper solid green)*

## Acknowledgements

R.P. Taylor is a Cottrell Scholar of the Research Corporation. B. Van Dusen was funded for this project by the M.J. Murdock Foundation, Partners in Science program. We thank M.S. Fairbanks for many fruitful discussions.